\documentclass[preprint]{aastex}
\newcommand{\HI}{H{\sc i}{ }}
\newcommand{\hi}{H{\sc i}}
\newcommand{\HII}{H{\sc~ii}{ }}
\newcommand{\kms}{km~s$^{-1}$}
\newcommand{\msol}{M$_{\odot}$}

\newcommand\ds{dSph}
\newcommand\di{dIrr}

\begin{document}

\title{The \HI environment of the Sculptor dwarf spheroidal galaxy}

\author{Antoine Bouchard}
\affil{D\'epartement de physique and Observatoire du mont M\'egantic, 
Universit\'e de Montr\'eal, C.P. 6128, Succ.
Centre-ville, Montr\'eal, Qu\'ebec, Canada H3C 3J7, and \\
Australia Telescope National Facility, PO Box 76, Epping, NSW 1710, Australia}
\email{bouchard@astro.umontreal.ca}

\author{Claude Carignan}
\affil{D\'epartement de physique and Observatoire du mont M\'egantic, 
Universit\'e de Montr\'eal, C.P. 6128, Succ.
Centre-ville, Montr\'eal, Qu\'ebec, Canada H3C 3J7} 
\email{carignan@astro.umontreal.ca}

\author{Sergey Mashchenko}
\affil{D\'epartement de physique and Observatoire du mont M\'egantic, 
Universit\'e de Montr\'eal, C.P. 6128, Succ. 
Centre-ville, Montr\'eal, Qu\'ebec, Canada H3C 3J7}
\email{syam@astro.umontreal.ca}

\begin{abstract}

New observations of the neutral hydrogen (\hi) in and around the line of sight
of the Sculptor dwarf spheroidal (dSph) are presented. The data obtained with
the single-dish Parkes telescope cover a large area of $7\degr\times 7\degr$ in
the direction of the dwarf, and have resolutions of $15\farcm 5\times
1.12$~km~s$^{-1}$. The Australia Telescope Compact Array was used to map a
smaller area of $2\fdg 2\times 2\fdg 2$ centered on the direction of the dwarf
with higher resolutions ($350\arcsec\times 140\arcsec\times 1.65$~km~s$^{-1}$).
Many \HI structures having velocities outside the range of the normal Galactic
disk velocities were detected, including the two Sculptor clouds (northeast and
southwest) of Carignan et al. (1998, C98). The present study shows the total
extent of the C98 clouds.  We derived heliocentric radial velocities for the NE
and SW clouds of $100.2\pm 0.9$~km~s$^{-1}$ and $105.1\pm 0.3$~km~s$^{-1}$,
respectively. The intensity-weighted mean \HI velocity for both clouds is
$104.1\pm 0.4$~km~s$^{-1}$.

Three different hypotheses concerning the association of the C98 Sculptor clouds
were considered. The case for the clouds belonging to the Sculptor group of
galaxies is found to be inconsistent with the observational data. The
probability of the C98 Sculptor clouds to be Milky Way features at anomalous
velocities (HVCs) superimposed by chance on the Sculptor dSph is estimated to be
less than 2\%. The third hypothesis assumes that the clouds are physically
associated with the Sculptor dSph, and is supported by the following evidences:
(a) the radial velocities for both clouds are very close to the optical velocity
of the Sculptor dSph ($\Delta V=4\pm 3$~km~s$^{-1}$).  (b) 88\% of the total \HI
flux is contained within the optical radius of the galaxy, and (c) the clouds
are located symmetrically relative to the center of the Sculptor dSph. Arguments
are presented that the C98 Sculptor clouds are still gravitationally bound to
the dwarf galaxy, and are part of its interstellar medium. The mass of each
cloud is $(4.1\pm 0.2)\times 10^4$~M$_\odot$ (NE cloud) and $(1.93\pm
0.02)\times 10^5$~M$_\odot$ (SW cloud) at the Sculptor dSph distance (79 kpc).

\end{abstract}


\section{Introduction}

Dwarf galaxies are the most numerous type of galaxies in the Universe. Recently,
the interest for the lowest luminosity dwarfs has been renewed, both on the
observational and on the theoretical sides, in part because these stellar
systems have the spatial scale at which the predictions of the presently favored
$\Lambda$CDM cosmology have the largest discrepancies with observations.  Most
notably, $\Lambda$CDM simulations overpredict the number of low mass galaxies in
the Milky Way and M31 halos \citep{moore99}. Another difficulty is the cuspy
density profile of simulated halos, which is at odds with the almost flat cores
observed in the inner few kiloparsecs of disk galaxies \citep{deblok01, bac01}.

Low luminosity dwarfs in the Local Group are classified as dwarf irregulars
(dIrr), dwarf spheroidals (dSph), or intermediate type dwarfs (dIrr/dSph)
\citep{mateo98}. \di{} galaxies are gas rich and the \HI is often seen in rings
surrounding the optical center \citep{carignan98, young97b} or bubble like
structures (LeoA, \citealt{young96}) and their complex Star Formation History
(SFH) can easily be explained. For \ds{} galaxies, the data are harder to
explain. The ISM is required to support the complex SFH \citep{grebel98} of
these objects and until now, very few of them are actualy known to contain \HI.
This is what led \citet{blitz2000} to believe ``that all of the LG dwarf
galaxies have had loosely bound \HI envelopes'' thus implicitly making a link
between High Velocity Clouds, \ds{}, and \di{}. 

In the Local Group, all dwarf galaxies less luminous than M$_{\rm V}\simeq
-15^{\rm m}$ and located within 250~kpc from a giant spiral (Milky Way or M31)
are dSphs. All dIrr galaxies in the same luminosity range are isolated systems.
This segregation of the dwarfs into dSphs and dIrrs according to their distance
from giant spirals suggests that some (or many) environmental factors are at
work. The most popular mechanisms discussed in the literature are total removal
of the ISM by either tidal or ram pressure stripping \citep[e.g.][]{blitz2000}.
However, the total ISM removal cannot explain the puzzling situation with many
dSphs, which have formed stars in the last 1-2~Gyrs, but have no \HI gas around
them which could have fueled the star formation. 

\citet*[ hereafter MCB]{syam2003} proposed a possible solution for this puzzle.
They showed that far ultraviolet (FUV) flux from spirals like the Milky Way and
M31 can be strong enough to keep the ISM of satellite dSphs in a fully
photoionized state for prolonged periods of time, and by assuming that many
dSphs are massive enough (with the virial temperature $T_{\rm vir}\gtrsim
10^4$~K) to keep the photoionized gas gravitationally bound. Only during
relatively short intervals of time, when the dSph moving along its orbit around
the host galaxy is crossing the shadow produced by the \HI disk of the host, can
the ISM recombine and potentially form stars.

Any evidence for a physical association of \HI gas with a low luminosity dSph
galaxy would be crucial for understanding the factors governing the evolution of
these systems.  One of the most promising objects in this regard is the Sculptor
dSph. \citet[hereafter C98]{carignan98b} detected two \HI clouds in the vicinity
of the dwarf, with radial velocities very similar to the optical velocity of the
Sculptor dSph. Their observations of a $\sim 0\fdg 5$ field covered only a small
portion of the stellar body of the dwarf which has a major axis of $\sim 2\fdg
5$ \citep{irwin95}. This prevented the authors from giving a conclusive evidence
for a physical association of the gas with the Sculptor dSph.  New \HI
observations with a much larger spatial coverage were required to see the full
extent of the clouds, and to make sure that these clouds were not Galactic HVCs.

Here we present new observations made with the Australia Telescope Compact Array
(ATCA) and the Parkes single-dish telescope covering large areas around the
Sculptor dwarf -- $7\degr \times 7\degr$ for Parkes and $2\degr \times 2\degr$
for the ATCA. We analyze all \HI structures detected in this area, and argue
that the new data strengthen the case for the physical association of the two
C98 clouds with the dSph while the others are most likely unrelated to the
Sculptor \ds. By combining our new \HI data with the data of \citet*{weiner2001}
on H$_\alpha$ emission from the Sculptor clouds, we also give evidence in
support of the radiation harassment scenario of MCB.

\section{Observations of \HI in the direction of the Sculptor dSph}
\label{Obs}

\subsection{Previous observations}

\begin{table}[tbp]
\caption{The Sculptor dSph physical parameters.}
\label{sclparam}
\begin{center}
\begin{tabular}{l l}
\tableline
\tableline
Parameter					&	Value 	\\
\tableline
RA (J2000)                                                   &  $1^{\rm h}00^{\rm m}09\fs 4$	\\
DEC (J2000)                                                  &  $-33\degr 42\arcmin 33\arcsec$	\\
Galactic longitude, l                                        &  $287\fdg 53$	\\
Galactic latitude, b                                         &  $-83\fdg 16$	\\
Heliocentric distance\tablenotemark{a}                       &  $79\pm 4$~kpc		\\
Isophotal major axis\tablenotemark{b}, $D_{25}$              &  $40\arcmin$	\\
Core radius\tablenotemark{c}, $r_c$	                     &  $5\farcm 8\pm 1\farcm 6$	\\
Tidal radius\tablenotemark{c}, $r_t$                         &  $76\farcm 5\pm 5\farcm 0$	\\
Major-axis position angle\tablenotemark{c}                   &  $99\degr\pm 1\degr$	\\
Optical radial velocity\tablenotemark{a}, V$_{\odot}^{opt}$  &  $108\pm 3$~\kms  \\
Proper motion velocity\tablenotemark{d}, V$_{prop}$          &  $210\pm 125$~\kms 	\\
Proper motion position angle\tablenotemark{d}                &  $40\degr \pm 24\degr$	\\

\tableline
\end{tabular}
\tablenotetext{a}{\citealt{mateo98}}
\tablenotetext{b}{\citealt{devaucouleurs91}}
\tablenotetext{c}{\citealt{irwin95}}
\tablenotetext{d}{\citealt{schweitzer95}}
\end{center}
\end{table}

C98 made two sets of observations. First, they used the Parkes single dish
radio-telescope and pointed it directly at the center of the galaxy. They used
the source PKS1934-638 as a flux calibrator and they had a velocity resolution
of 1.65~\kms. Their 100 minutes integration on source resulted in a detection
limit of 100~\msol. The spectrum they obtained revealed $\sim 10^{4}$~\msol{} of
H{\sc i}, at a velocity of V$_{\odot}$ = 112~km~s$^{-1}$. But the
$\sim$15\arcmin{} beam did not cover all the optical extent of the Sculptor dSph
(which has a tidal radius of $\sim 1\fdg 3$, see Table~\ref{sclparam}) and it
was not large enough to get a complete sampling of what is believed to be the
gas associated with the dwarf.

Their second set of observations was obtained with the ATCA using the array
configuration 210 which resulted in a spatial resolution of 240\arcsec{} and a
velocity resolution of 1.65~\kms. This yielded an \HI mass of
3.0$\times10^{4}$~\msol{}, at a velocity of $102\pm 5$~km~s$^{-1}$.  But
inevitable short spacing effects appeared in the data which made the authors
believe that they missed a non-negligible fraction of the \HI. Moreover a large
fraction of the detected \HI lied close to the half-power beam width
(FWHM$=33\arcmin$).

\subsection{New Parkes data}

The 64~m Parkes radio-telescope was used, equipped with a 21~cm multibeam
receiver in a narrowband mode. The total bandwidth of 8~MHz was divided in 2048
channels, resulting in a velocity increment of 0.82~\kms{} between each channel
of the cube for a total velocity coverage of 1600~\kms{} centered on
V$_{\odot}$~=~-200~\kms. A region of $7\degr\times 7\degr$ was mapped around the
Sculptor dSph at a beam resolution of $15.5\arcmin$ and a velocity resolution of
1.12~\kms, using frequency switching observations.

The reduction pipeline provided with the AIPS++ package was used to calibrate
the data (task "Livedata") and to grid and form the cube (task "Gridzilla").
Both tasks are part of the standard HIPASS data acquisition system
\citep{barnes2001}. For the bandpass calibration, the data were Hanning smoothed
and the median value of the baseline was subtracted.

\subsection{New ATCA data}

The ATCA is located in Narrabri, Australia. It consists of six 22 meters
antennae on an East-West track. This aperture synthesis array was used in the
0.210 configuration, with baselines ranging from 31 to 214 meters, giving the
maximum possible weight to large structures. 

The data cube is a 64 pointing mosaic centered on a frequency of 1420~MHz. The
array configuration, along with a total bandpass of 8~MHz divided in 1024
channels resulted in physical resolutions of $350\arcsec\times 140\arcsec\times
1.65$~\kms. The source PKS1934-638 was used as a flux calibrator while
PKS0008-421 was the phase calibrator. The Miriad package was used to edit and
calibrate the data.

\subsection{Data Analysis}

\begin{figure}[tbh]
\begin{center}
\includegraphics[width=0.45\textwidth]{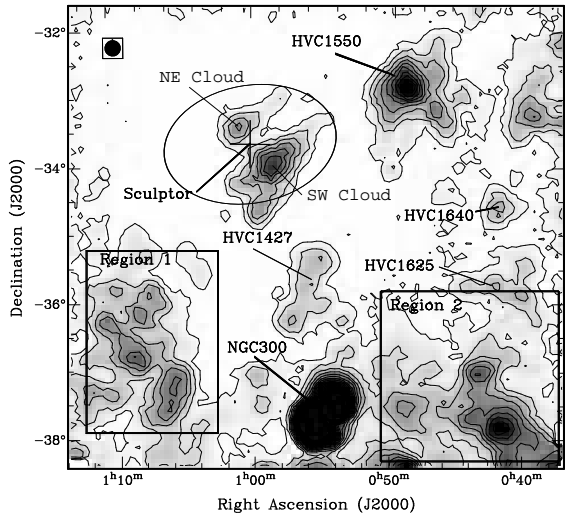}
\caption{Parkes narrowband integrated \HI map. The contour levels are (1, 5, 10,
15, \dots, 40)$\times 10^{18}$~cm$^{-2}$.  We show the locations of some HVCs
from the catalog of \citet{putman2002}, and define two regions of comparable
velocities --- Region 1 and Region 2.  The ellipse shows the optical extent of
the Sculptor dSph \citep{irwin95}.  The cross marks the location of the optical
center of Sculptor.}\label{pksmom0}
\end{center}
\end{figure}

Figure~\ref{pksmom0} shows a wide field image taken with the Parkes
radiotelecope equiped with the narrowband multibeam receiver.  The region in the
direction of the Sculptor \ds{} is heavily populated with \HI features. Aside
from the Sculptor (NE and SW) clouds, only one \HI structure from
Figure~\ref{pksmom0} has been previously associated with an optical object ---
the spiral galaxy NGC~300 from the Sculptor group. Many other features are
catalogued as HVCs by \citet{putman2002}. The radial velocities for each feature
marked in Figure~\ref{pksmom0} are listed in Table~\ref{hvcparam}.

\begin{figure}[tbh]
\begin{center}
\includegraphics[width=0.45\textwidth]{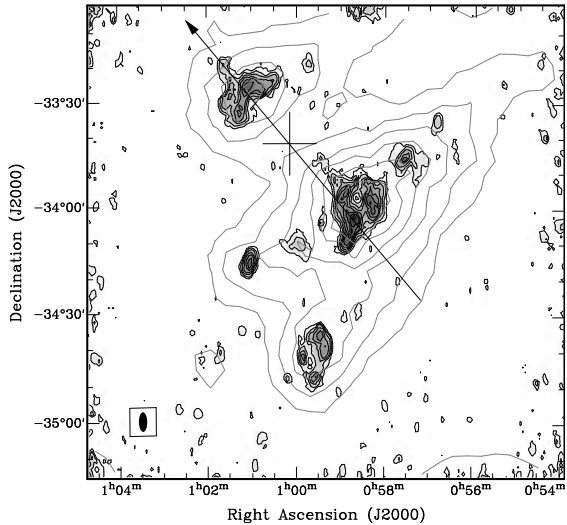}
\caption{ATCA integrated \HI (grey-scale image and black contours) overlayed
with the Parkes data from Figure~\ref{pksmom0} (grey contours). The contours for
the ATCA data have levels (5, 10, 15, \dots, 40)$\times 10^{18}$~cm$^{-2}$. The
arrow shows the direction of the proper motion of the Sculptor dSph from
\citet{schweitzer95}.}\label{camom0}
\end{center}
\end{figure}

\begin{figure}[tbh]
\begin{center}
\includegraphics[width=0.45\textwidth]{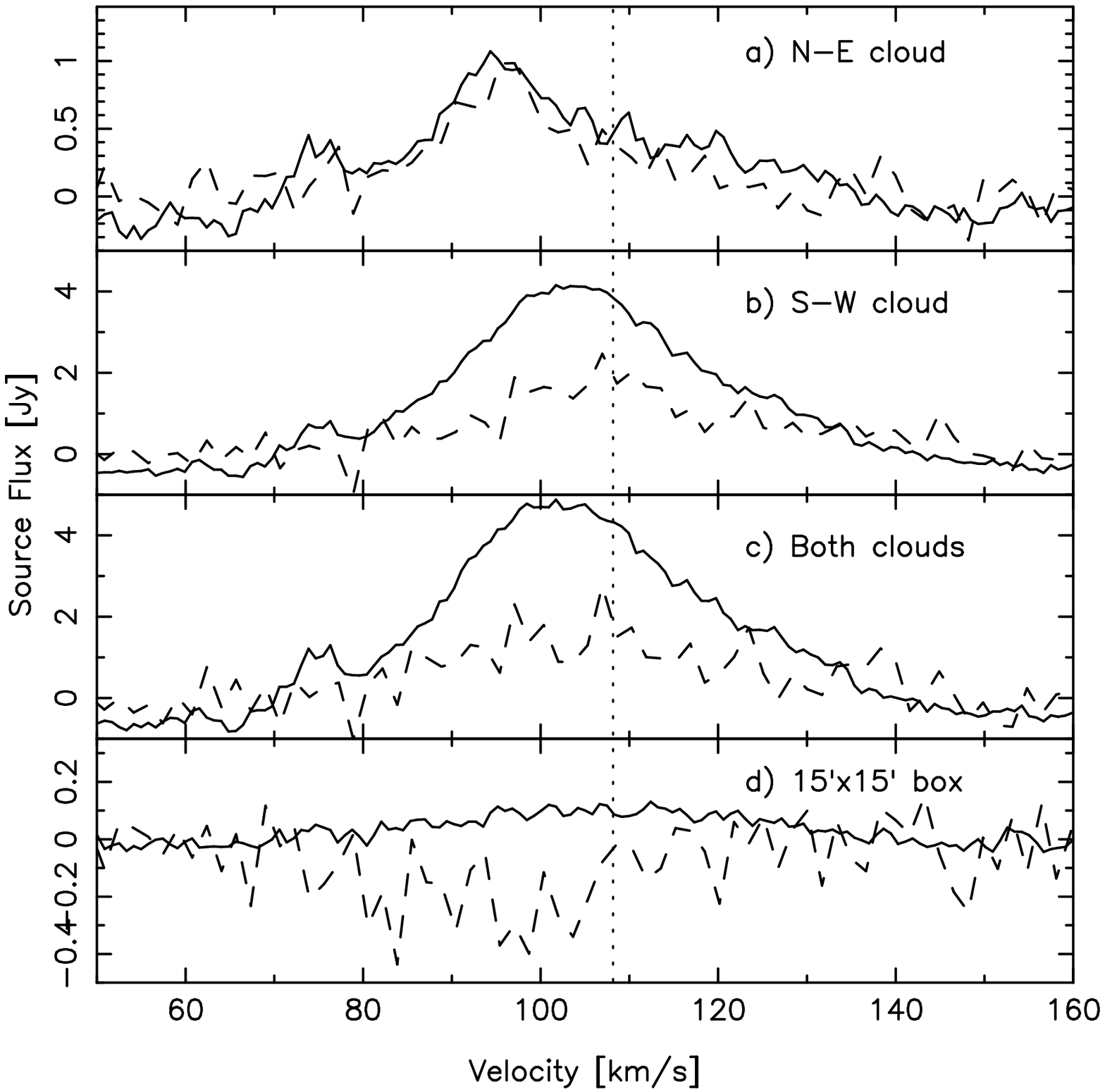}
\caption{Four spectra, taken both from the Parkes (solid line) and ATCA (dashed
line) data sets. Panels a) and b) show the spectra integrated over the
North-East and South-West clouds, respectively, panel c) shows the total
spectrum, and panel d) shows the spectrum for a 15\arcmin{} by 15\arcmin{} box
centered on the dwarf. The vertical doted line corresponds to the optical
radial velocity of the Sculptor dSph.  }\label{spectre} 
\end{center}
\end{figure}

This study will concentrate on the two \HI features closest to the optical
center of the Sculptor dSph --- the NE and SW clouds (see Figure~\ref{pksmom0}).
A close-up view of the Sculptor clouds is shown in Figure~\ref{camom0}. This is
a higher spatial resolution image taken with the ATCA interferometer. The
integrated spectra for the clouds are shown in Figure~\ref{spectre}. The
velocity information extracted from the spectra is summarized in
Table~\ref{tabspectre}. The table also contains \HI mass estimates, based on the
assumption that the clouds are located at the distance of the Sculptor dSph
(79~kpc, see Table~\ref{sclparam}).

\begin{table}[tbh]
\caption{Properties of the \HI features of Figure \ref{pksmom0}.}
\label{hvcparam}
\begin{center}
\renewcommand{\thefootnote}{\thempfootnote}
\renewcommand{\footnoterule}{\rule{1cm}{0cm}\vspace{-0.4cm}}
\begin{tabular}{l c c}
\tableline
\tableline
		& V$_\odot$ &	$\sigma_{max}$ \\
		& (\kms)	& (\kms) \\
\tableline
NE cloud (HVC 1321)	&	100 &	6.0\\
SW cloud (HVC 1353)	&	105 &	9.9\\
NGC 300			&	145 &	37.6\\
HVC 1427		&	135 &	7.9\\
HVC 1550		&	117 &	11.0\\
HVC 1625		&	169 &	21.5\\
HVC 1640		&	175 &	13.4\\
Region 1 (broad region)	&	121 &	9.0\\
Region 1 (artifact)	&	121 &	21.0\\
Region 2		&	40  &	4.4\\
\tableline
\end{tabular}
\end{center}
\end{table}

\begin{table}[tbh]
\caption{List of all HVCs within 10\degr{} of the Sculptor \ds{} line of sight
\citep{putman2002} and within the 80 \kms{} to 210 \kms{} velocity range.}
\label{hvclistin}
\begin{center}
\renewcommand{\thefootnote}{\thempfootnote}
\renewcommand{\footnoterule}{\rule{1cm}{0cm}\vspace{-0.4cm}}
\begin{tabular}{l c c c c c}
\tableline
\tableline
Catalog & RA      & DEC     & $\Delta \alpha$  & V$_{LSR}$ & $\Delta V$ \\
designation  & (J2000) & (J2000) & & (\kms)  & (\kms)     \\
\tableline
HVC 1520 & 0$^h$ 50.3$^m$ & -34$\degr$  4.0$^{\prime}$ & 2.08$\degr$ &   140 &     39 \\
HVC 1427 & 0$^h$ 56.1$^m$ & -35$\degr$ 53.0$^{\prime}$ & 2.33$\degr$ &   126 &     25 \\
HVC 1550 & 0$^h$ 48.8$^m$ & -32$\degr$ 59.0$^{\prime}$ & 2.48$\degr$ &   113 &     12 \\
HVC 1634 & 0$^h$ 46.8$^m$ & -31$\degr$ 20.0$^{\prime}$ & 3.68$\degr$ &   138 &     37 \\
HVC 1640 & 0$^h$ 42.4$^m$ & -34$\degr$ 40.0$^{\prime}$ & 3.79$\degr$ &   160 &     59 \\
HVC 1547 & 0$^h$ 47.2$^m$ & -37$\degr$ 11.0$^{\prime}$ & 4.36$\degr$ &   141 &     40 \\
HVC 1625 & 0$^h$ 42.0$^m$ & -36$\degr$  4.0$^{\prime}$ & 4.41$\degr$ &   160 &     59 \\
HVC 1540 & 0$^h$ 47.3$^m$ & -38$\degr$ 20.0$^{\prime}$ & 5.30$\degr$ &   164 &     63 \\
HVC 1462 & 0$^h$ 55.7$^m$ & -39$\degr$ 22.0$^{\prime}$ & 5.73$\degr$ &   144 &     43 \\
HVC 1723 & 0$^h$ 31.7$^m$ & -36$\degr$ 29.0$^{\prime}$ & 6.45$\degr$ &   202 &    101 \\
HVC 912 & 1$^h$ 31.3$^m$ & -34$\degr$ 53.0$^{\prime}$ & 6.54$\degr$ &   107 &      6 \\
HVC 1791 & 0$^h$ 26.6$^m$ & -35$\degr$ 52.0$^{\prime}$ & 7.22$\degr$ &   146 &     45 \\
:HVC 1569 & 0$^h$ 43.1$^m$ & -40$\degr$  8.0$^{\prime}$ & 7.27$\degr$ &   186 &     85 \\
:HVC 1616 & 0$^h$ 38.6$^m$ & -39$\degr$ 45.0$^{\prime}$ & 7.42$\degr$ &   158 &     57 \\
HVC 1908 & 0$^h$ 23.0$^m$ & -32$\degr$ 49.0$^{\prime}$ & 7.82$\degr$ &   129 &     28 \\
HVC 1825 & 0$^h$ 22.0$^m$ & -35$\degr$ 40.0$^{\prime}$ & 8.08$\degr$ &   171 &     70 \\
HVC 1901 & 0$^h$ 20.5$^m$ & -33$\degr$ 37.0$^{\prime}$ & 8.25$\degr$ &   198 &     97 \\
HVC 3 & 0$^h$ 21.9$^m$ & -30$\degr$ 58.0$^{\prime}$ & 8.53$\degr$ &   109 &      8 \\
HVC 732 & 1$^h$ 42.4$^m$ & -32$\degr$ 22.0$^{\prime}$ & 8.95$\degr$ &   111 &     10 \\
HVC 1932 & 0$^h$ 15.7$^m$ & -33$\degr$  7.0$^{\prime}$ & 9.29$\degr$ &   146 &     45 \\
HVC 1280 & 1$^h$ 20.4$^m$ & -42$\degr$  7.0$^{\prime}$ & 9.30$\degr$ &   148 &     47 \\
:HVC 1614 & 0$^h$ 36.6$^m$ & -42$\degr$  4.0$^{\prime}$ & 9.56$\degr$ &   145 &     44 \\
HVC 1922 & 0$^h$ 13.6$^m$ & -33$\degr$ 53.0$^{\prime}$ & 9.67$\degr$ &   162 &     61 \\
:HVC 1651 & 0$^h$ 30.6$^m$ & -41$\degr$ 38.0$^{\prime}$ & 9.84$\degr$ &   171 &     70 \\
CHVC 1104 & 1$^h$ 35.5$^m$ & -40$\degr$ 38.0$^{\prime}$ & 9.86$\degr$ &   120 &     19 \\
HVC 1031 & 1$^h$ 41.0$^m$ & -39$\degr$ 23.0$^{\prime}$ & 9.96$\degr$ &   101 &      0 \\
\tableline
\end{tabular}
\end{center}
\end{table}

\begin{table}[tbh]
\caption{List of all HVCs within 10\degr{} of the Sculptor \ds{} line of sight
\citep{putman2002} and outside of the 80 \kms{} to 210 \kms{} velocity range.}
\label{hvclistout}
\begin{center}
\begin{footnotesize}
\renewcommand{\thefootnote}{\thempfootnote}
\renewcommand{\footnoterule}{\rule{1cm}{0cm}\vspace{-0.4cm}}
\begin{tabular}{l c c c c c}
\tableline
\tableline
Catalog & RA      & DEC     & $\Delta \alpha$  & V$_{LSR}$ & $\Delta V$ \\
designation  & (J2000) & (J2000) & & (\kms)  & (\kms)     \\
\tableline
HVC 1247 & 1$^h$  4.4$^m$ & -33$\degr$ 52.0$^{\prime}$ & 0.90$\degr$ &  -212 &    312 \\
HVC 1292 & 1$^h$  0.7$^m$ & -32$\degr$ 47.0$^{\prime}$ & 0.93$\degr$ &  -180 &    280 \\
CHVC 1387 & 0$^h$ 58.0$^m$ & -35$\degr$  1.0$^{\prime}$ & 1.38$\degr$ &  -243 &    343 \\
HVC 1175 & 1$^h$  5.4$^m$ & -32$\degr$ 52.0$^{\prime}$ & 1.38$\degr$ &  -140 &    240 \\
HVC 1461 & 0$^h$ 53.8$^m$ & -34$\degr$ 31.0$^{\prime}$ & 1.54$\degr$ &  -167 &    267 \\
HVC 1192 & 1$^h$  0.3$^m$ & -31$\degr$  2.0$^{\prime}$ & 2.68$\degr$ &  -182 &    282 \\
HVC 1385 & 0$^h$ 54.9$^m$ & -31$\degr$ 15.0$^{\prime}$ & 2.70$\degr$ &  -191 &    291 \\
HVC 1538 & 0$^h$ 49.8$^m$ & -31$\degr$ 54.0$^{\prime}$ & 2.83$\degr$ &  -170 &    270 \\
:HVC 1411 & 0$^h$ 57.9$^m$ & -36$\degr$ 57.0$^{\prime}$ & 3.27$\degr$ &  -205 &    305 \\
HVC 1624 & 0$^h$ 44.4$^m$ & -33$\degr$ 56.0$^{\prime}$ & 3.28$\degr$ &  -199 &    299 \\
HVC 1644 & 0$^h$ 45.5$^m$ & -32$\degr$  8.0$^{\prime}$ & 3.46$\degr$ &  -137 &    237 \\
HVC 1567 & 0$^h$ 49.3$^m$ & -30$\degr$ 50.0$^{\prime}$ & 3.68$\degr$ &  -191 &    291 \\
HVC 1251 & 0$^h$ 56.8$^m$ & -30$\degr$  4.0$^{\prime}$ & 3.71$\degr$ &  -143 &    243 \\
HVC 1719 & 0$^h$ 42.4$^m$ & -31$\degr$ 43.0$^{\prime}$ & 4.23$\degr$ &  -152 &    252 \\
HVC 1021 & 1$^h$ 21.4$^m$ & -34$\degr$ 54.0$^{\prime}$ & 4.55$\degr$ &   -88 &    188 \\
HVC 1706 & 0$^h$ 37.4$^m$ & -34$\degr$ 45.0$^{\prime}$ & 4.82$\degr$ &  -166 &    266 \\
HVC 1532 & 0$^h$ 51.0$^m$ & -28$\degr$ 42.0$^{\prime}$ & 5.38$\degr$ &  -202 &    302 \\
HVC 854 & 1$^h$ 28.6$^m$ & -33$\degr$ 23.0$^{\prime}$ & 5.93$\degr$ &  -121 &    221 \\
HVC 892 & 0$^h$ 54.2$^m$ & -27$\degr$ 43.0$^{\prime}$ & 6.13$\degr$ &  -212 &    312 \\
HVC 707 & 1$^h$ 23.6$^m$ & -30$\degr$  0.0$^{\prime}$ & 6.21$\degr$ &  -164 &    264 \\
HVC 1818 & 0$^h$ 30.0$^m$ & -33$\degr$ 43.0$^{\prime}$ & 6.27$\degr$ &  -172 &    272 \\
HVC 615 & 0$^h$ 59.6$^m$ & -27$\degr$ 22.0$^{\prime}$ & 6.34$\degr$ &  -256 &    356 \\
CHVC 1664 & 0$^h$ 35.7$^m$ & -37$\degr$ 52.0$^{\prime}$ & 6.47$\degr$ &   215 &    114 \\
HVC 1903 & 0$^h$ 31.0$^m$ & -31$\degr$ 31.0$^{\prime}$ & 6.52$\degr$ &  -192 &    292 \\
:HVC 879 & 1$^h$ 32.2$^m$ & -34$\degr$ 23.0$^{\prime}$ & 6.67$\degr$ &  -155 &    255 \\
HVC 941 & 1$^h$ 32.3$^m$ & -35$\degr$ 34.0$^{\prime}$ & 6.87$\degr$ &   -83 &    183 \\
HVC 1499 & 0$^h$ 51.2$^m$ & -40$\degr$ 31.0$^{\prime}$ & 7.04$\degr$ &  -110 &    210 \\
CHVC 630 & 1$^h$ 20.7$^m$ & -28$\degr$ 12.0$^{\prime}$ & 7.05$\degr$ &  -197 &    297 \\
HVC 764 & 1$^h$ 33.4$^m$ & -32$\degr$ 22.0$^{\prime}$ & 7.09$\degr$ &   -96 &    196 \\
HVC 698 & 1$^h$ 30.0$^m$ & -30$\degr$ 11.0$^{\prime}$ & 7.24$\degr$ &   -93 &    193 \\
HVC 588 & 1$^h$  1.9$^m$ & -26$\degr$ 28.0$^{\prime}$ & 7.25$\degr$ &  -268 &    368 \\
HVC 673 & 1$^h$ 28.4$^m$ & -29$\degr$ 28.0$^{\prime}$ & 7.36$\degr$ &  -118 &    218 \\
HVC 1777 & 0$^h$ 26.4$^m$ & -36$\degr$ 27.0$^{\prime}$ & 7.43$\degr$ &   255 &    154 \\
CHVC 542 & 0$^h$ 54.1$^m$ & -25$\degr$ 59.0$^{\prime}$ & 7.84$\degr$ &  -240 &    340 \\
HVC 1753 & 0$^h$ 25.9$^m$ & -37$\degr$ 31.0$^{\prime}$ & 7.93$\degr$ &  -130 &    230 \\
HVC 219 & 0$^h$ 36.2$^m$ & -27$\degr$ 40.0$^{\prime}$ & 7.94$\degr$ &  -168 &    268 \\
HVC 827 & 1$^h$ 41.8$^m$ & -34$\degr$ 32.0$^{\prime}$ & 8.65$\degr$ &   -97 &    197 \\
HVC 1592 & 0$^h$ 39.2$^m$ & -41$\degr$ 37.0$^{\prime}$ & 8.92$\degr$ &  -100 &    200 \\
HVC 1794 & 0$^h$ 20.1$^m$ & -37$\degr$ 38.0$^{\prime}$ & 9.03$\degr$ &  -131 &    231 \\
CHVC 1079 & 1$^h$ 35.5$^m$ & -39$\degr$ 36.0$^{\prime}$ & 9.21$\degr$ &  -112 &    212 \\
HVC 432 & 0$^h$ 38.2$^m$ & -25$\degr$ 42.0$^{\prime}$ & 9.32$\degr$ &  -140 &    240 \\
HVC 1852 & 0$^h$ 15.6$^m$ & -36$\degr$  5.0$^{\prime}$ & 9.43$\degr$ &  -124 &    224 \\
HVC 1556 & 0$^h$ 42.6$^m$ & -42$\degr$ 36.0$^{\prime}$ & 9.53$\degr$ &   -98 &    198 \\
HVC 1483 & 0$^h$ 53.6$^m$ & -43$\degr$ 22.0$^{\prime}$ & 9.74$\degr$ &   226 &    125 \\
HVC 1641 & 0$^h$ 32.0$^m$ & -41$\degr$ 56.0$^{\prime}$ & 9.92$\degr$ &   -94 &    194 \\
\tableline
\end{tabular}
\end{footnotesize}
\end{center}
\end{table}

\begin{table*}[tbh]
\caption{Data on the Sculptor \HI clouds. The \HI masses assume that the clouds
are at the distance of the Sculptor \ds.}
\label{tabspectre}
\begin{center}
\begin{tabular}{l c c c c}
\tableline
\tableline
	& \multicolumn{2}{c}{Parkes} & \multicolumn{2}{c}{ATCA}\\
	& Velocity & \HI mass & Velocity & \HI mass \\
	& (\kms) & ($\times$10$^4$\msol~[D/79~kpc]$^2$) & (\kms) & ($\times$10$^4$\msol~[D/79~kpc]$^2$)\\
\tableline
North East & 100.2$\pm$0.9 & 4.07$\pm$0.18 & 99.7$\pm$1.7 &  2.84$\pm$0.23\\
South West & 105.1$\pm$0.3 & 19.33$\pm$0.24 & 110.2$\pm$1.6 & 8.67$\pm$0.68\\
Central $15\arcmin\times 15\arcmin$ & 106.3$\pm$0.7 & 0.61$\pm$0.02 & \nodata & (-1.13$\pm$0.27)\tablenotemark{a} \\
Total & 104.1$\pm$0.4 & 23.35$\pm$0.32 & 109.3$\pm$2.0 & 8.68$\pm$0.83 \\
\tableline
\end{tabular}
\tablenotetext{a}{See text for details}
\end{center}
\end{table*}

Parts of the Sculptor clouds were observed by C98. In our present study, we can
see the full extent of the clouds.  They have larger sizes and masses
(Table~\ref{tabspectre}) than those reported by C98, but appear to be almost
completely contained within the optical extent of the Sculptor dSph (88\% of the
total flux). The existence of the third (central) cloud of C98 with low
detection significance is not confirmed with our new data, which is evident from
the spectrum of the central $15\arcmin \times 15\arcmin$ area shown in
Figure~\ref{spectre}d and the corresponding parameters given in
Table~\ref{tabspectre}.  This central region has a slightly negative integral
flux (resulted in a negative mass in Table~\ref{tabspectre}). This is caused by
an imperfect continuum subtraction performed on the image. The signal does not
significantly come out of the noise (Figure~\ref{spectre}d).

The lack of short spacings in the ATCA data is most obvious in panels b) and c)
of Figure~\ref{spectre}. The 210 configuration of the array used for the
observations is not sensitive to scales larger than $\sim 30\arcmin$, resulting
in the mass of the SW cloud being significantly underestimated
(Table~\ref{tabspectre}). The NE cloud with a size $\sim 22\arcmin$ has less
suffered from this effect.

\begin{figure}[tbh]
\begin{center}
\includegraphics[height=0.23\textwidth, width=0.46\textwidth]{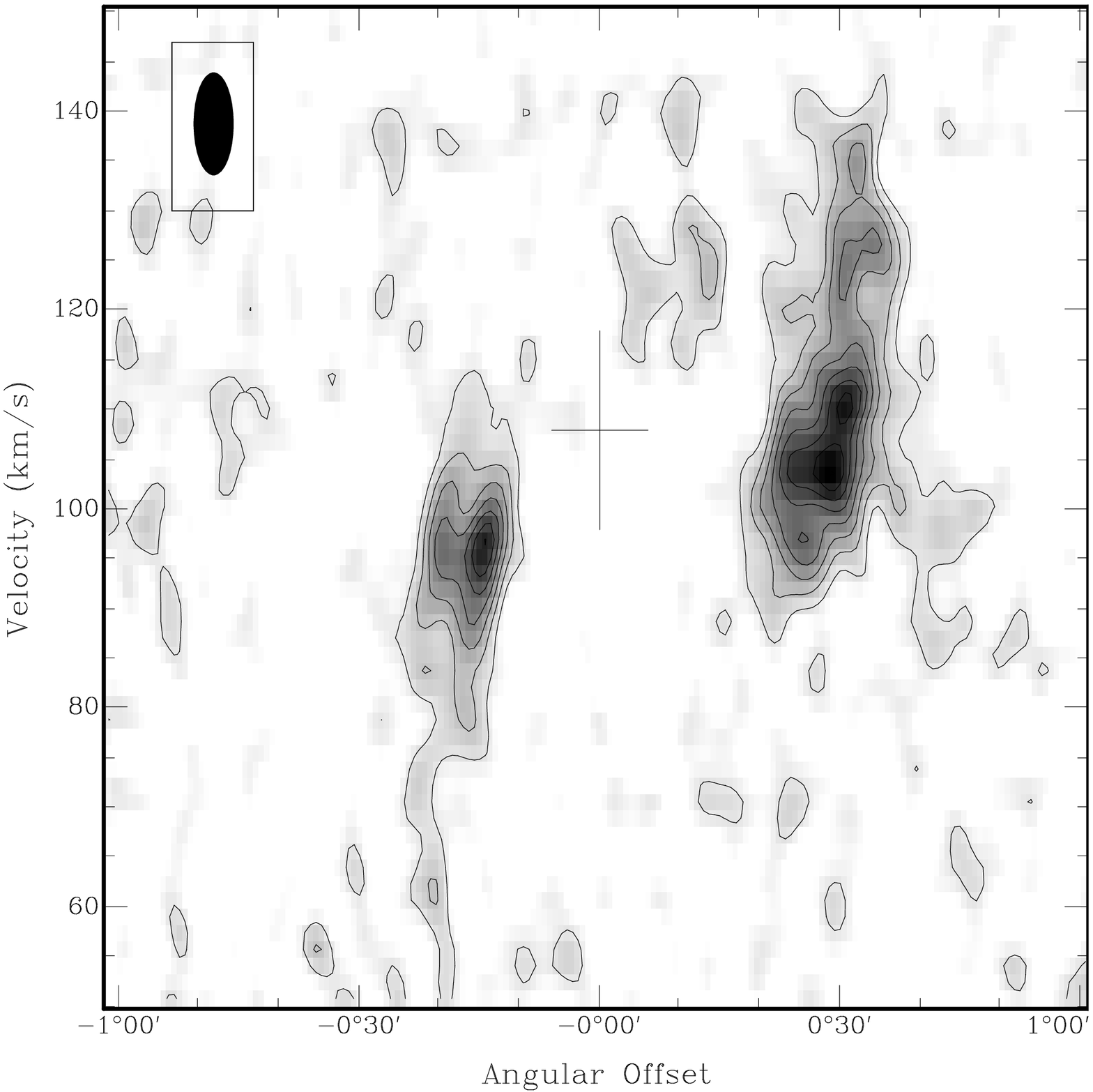}
\includegraphics[height=0.23\textwidth, width=0.46\textwidth]{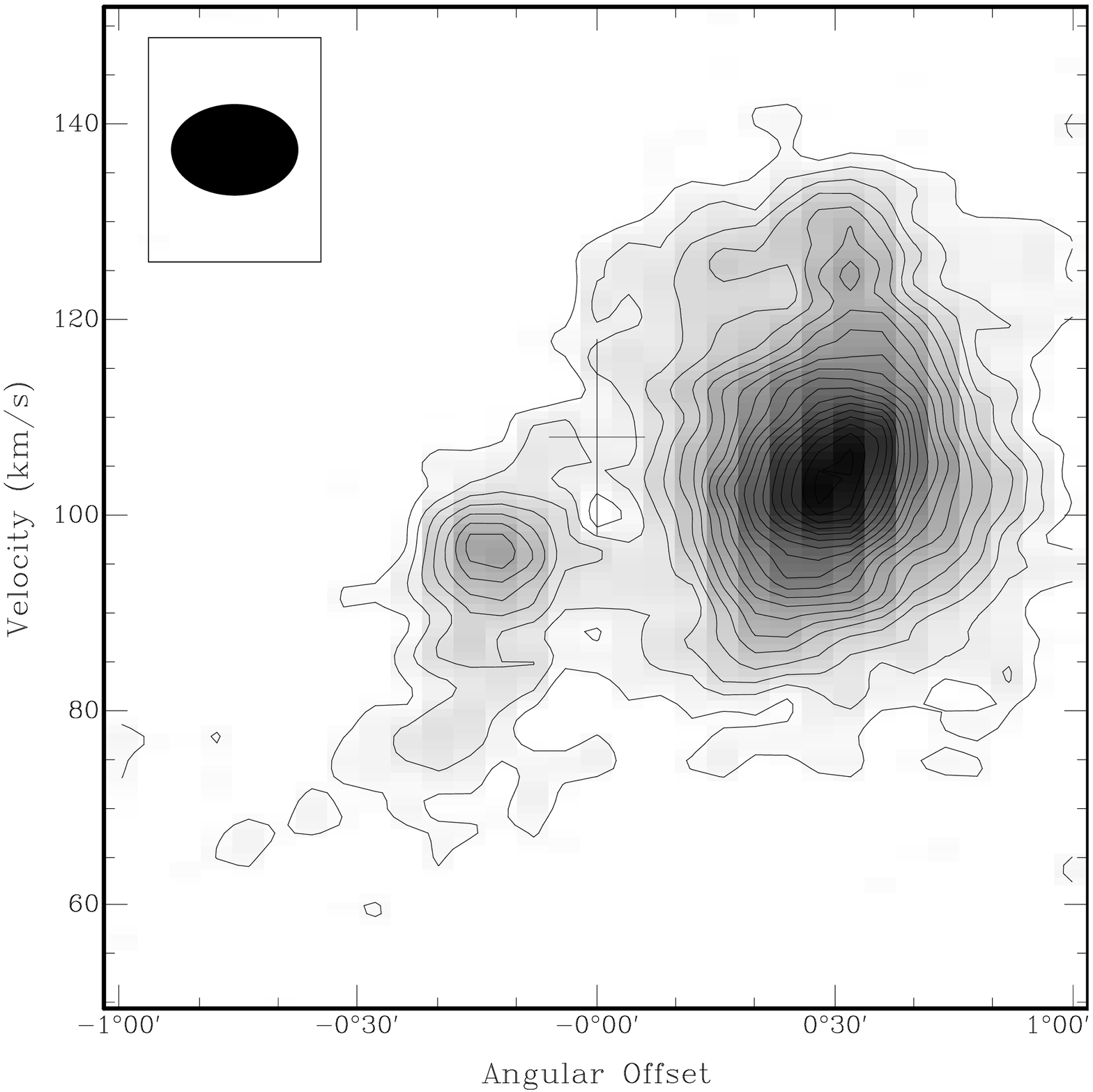}
\caption{A cut through the ATCA (top panel) and Parkes narrowband (bottom panel)
data sets. The slice is 12\arcmin{} wide in both cubes and is aligned with the
galaxy's proper motion (position angle of 40$\degr$, see Figure~\ref{camom0}).
The ATCA data have been smoothed in velocity to 11.41 \kms{} and have contour
levels from 10 to 70 mJy by steps of 10 mJy. The narrowband data are smoothed in
velocity to 10.66 \kms{} and the levels are from 2.5 to 50 mJy at every 2.5
mJy.}\label{lv} 
\end{center}
\end{figure}

Figure~\ref{lv} shows a slice along the velocity axis through both the ATCA and
the Parkes data sets. The slice passes through the center of the Sculptor dSph
along the proper motion direction (arrow in Figure~\ref{camom0}). We clearly see
the NE cloud (on the left side) and the SW cloud (on the right side). The cross
marks the position of the galactic center and the velocity of the dwarf. It is
clear from this figure that both clouds are distributed symmetrically relative
to the dwarf's center and have similar velocities, though the velocity of the NE
cloud tends to be slightly less than that of the SW cloud.

\begin{figure}[tbh]
\begin{center}
\includegraphics[width=0.45\textwidth]{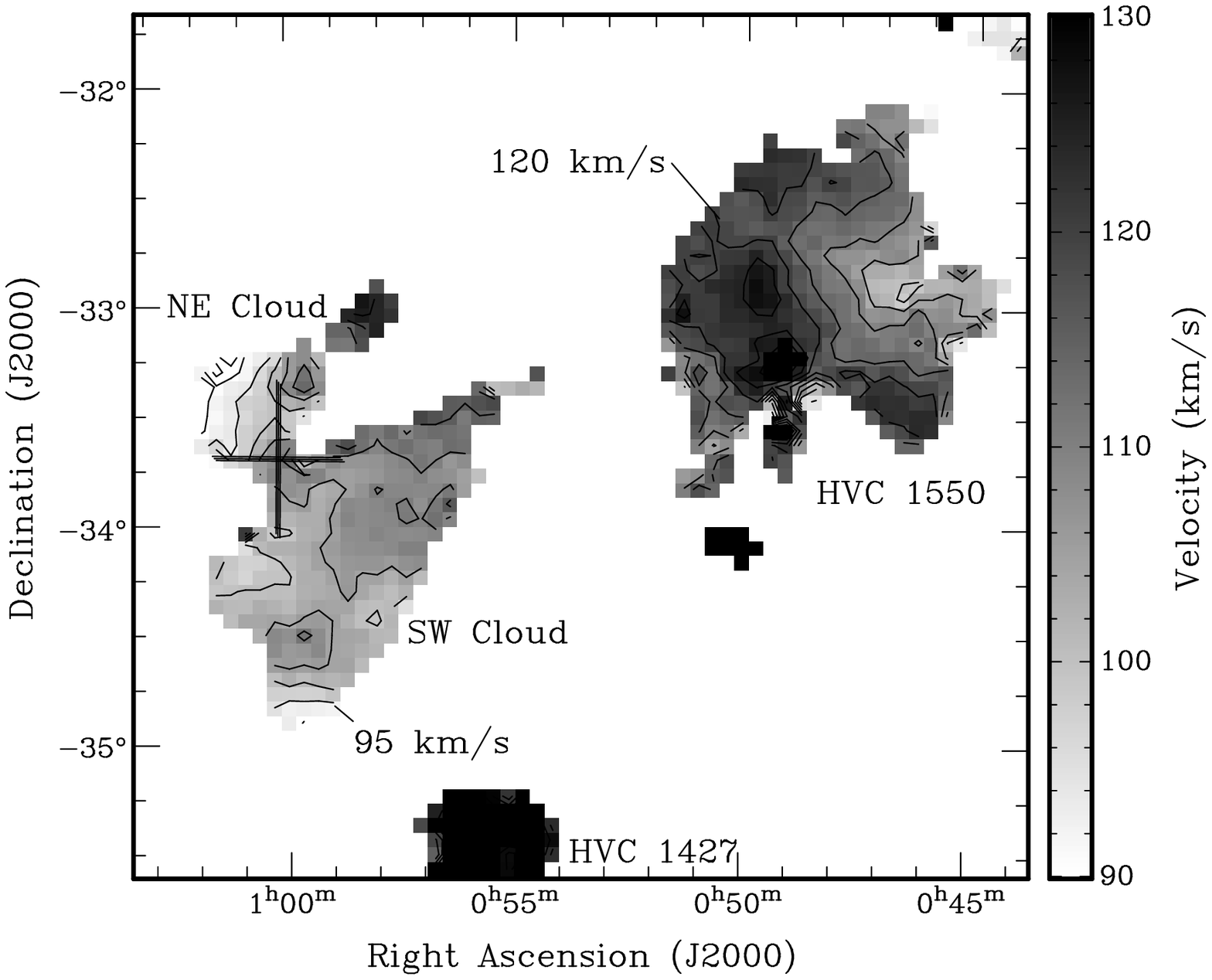}
\caption{The velocity map for the Parkes narrowband data. The contours are
separated by 5 \kms, ranging from 90 to 130 \kms. Only the closest clouds to the
Sculptor \ds{}, in both velocity and position on the sky, have been included in
this figure.  The wide-field nature of this picture and the high discrepancy in
the velocities of the clouds would make the full 7$\degr \times 7\degr$ Figure
irrelevant to the discussion and velocity gradients
undistinguishable.}\label{pksmom1}
\end{center}
\end{figure}

\begin{figure}[tbh]
\begin{center}
\includegraphics[width=0.45\textwidth]{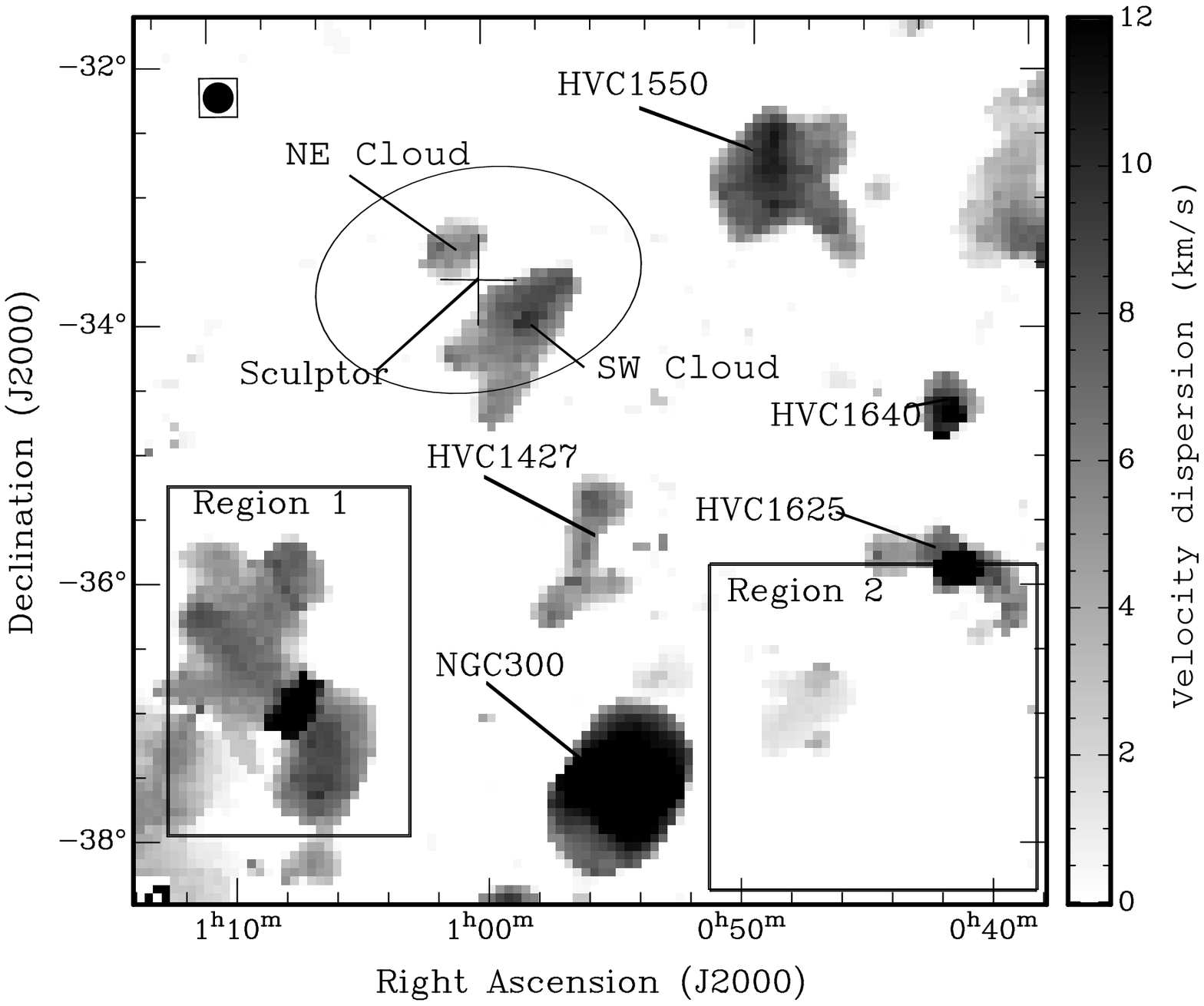}
\caption{Velocity dispersions from the Parkes dataset.}\label{pksmom2}
\end{center}
\end{figure}

Velocity information can also be found in Figure~\ref{pksmom1}. This velocity
map from the Parkes data shows the two main Sculptor clouds and a third one (HVC
1550) close to the dwarf.  Figure~\ref{pksmom2} shows the dispersion of the \HI
gas of the Sculptor dSph surroundings. The maximum dispersion values of every
feature identified in Figure~\ref{pksmom0} are listed in Table~\ref{hvcparam}.

\section{Discussion}
\subsection{Physical association of the Sculptor clouds}

The quality of our present data on the \HI distribution in the line of sight of
the Sculptor dSph gives us an opportunity to address the issue of the physical
association of \HI gas with a dSph in a more conclusive manner than any previous
study. In the past, the attempts to associate nearby \HI gas with dSph galaxies
have been hampered by some of the following factors: lack of spatial coverage
(C98), low angular and velocity resolutions \citep{blitz2000}, and lack of
stellar radial velocity information \citep*{oosterloo96}.  In this study, we
achieve both sufficiently large spatial coverage ($7\degr\times 7\degr$, with
the area covered being 14 times larger than the area of the Sculptor stellar
body) and high angular and velocity resolutions ($5\farcm8\times 2\farcm3\times
$1.65~km~s$^{-1}$) by combining the single-dish Parkes with intereferometric
ATCA observations. In addition, the stellar radial velocity of the Sculptor dSph
is now known to a high accuracy (Table~\ref{sclparam}).

The locations and radial velocities of the NE and SW clouds leave us with only two
possible alternatives in regards to the physical association with the Sculptor
dSph: the case of a larger distance (the clouds belong to the Sculptor group of
galaxies), and the case of a smaller distance (the clouds are Galactic HVCs). We
will consider both possibilities in turn.

At the distances of the Sculptor group members ($1.7-4.4$~Mpc,
\citealt{jerjen98}) the clouds would have very large size ($\sim 25-60$~kpc) and
total mass ($[1-5]\times 10^8$~M$_\odot$). Isolated intergalactic \HI clouds of
this size are not known. Moreover, the H$_\alpha$ flux of 0.22~R, detected in
the densest part of the SW cloud (see Section~\ref{nature}), is too large to be
caused by the metagalactic ionizing radiation background, and is most naturally
explained by the impact of the LyC radiation from the Milky Way, implying a
small distance to the clouds of $\lesssim 100$~kpc.  The clouds also cannot
belong to a galaxy located at the distance of the Sculptor group: a gas-rich
galaxy of this size would be a disk galaxy, which is at odds with both the
irregular shape of the clouds and the absence of a clear rotation signature. We
conclude that the Sculptor group hypothesis appears to be inconsistent with the
available data.

The second alternative --- the SW and NE clouds being HVCs --- is more difficult
to discard.  HVCs are clouds that do not fit any simple galactic rotation
models.  These clouds are often associated in large HVC complexes that are
believed to be of similar origin. One of these complexes, the Magellanic Stream,
has many components near the Sculptor dSph, but its velocity in this direction,
$\sim -60$~\kms{} \citep{mathewson84}, makes it a very unlikely association.

One should also realize that the majority of the HVCs is not part of an HVC
complex. What is seen in Figure~\ref{pksmom0} could be independent HVCs which
happened to be in the line of sight of the Sculptor dSph and have radial
velocities very close to the radial velocity of the dwarf. We estimated the
probability of such an event in the following manner. As can be seen in Table
\ref{hvclistin}, the number $N$ of HVCs from the catalog of \citet{putman2002}
located within $R_0=10\degr$ from the Sculptor dSph and having heliocentric
radial velocities within the interval $V=+80 \dots +210$~km~s$^{-1}$ is $N=26$
(excluding the Sculptor clouds). We considered the NE and SW clouds to be one
cloud. The center of mass of this cloud is located at the distance $R_1=0\fdg
318$ from the center of the Sculptor dSph, and has a radial velocity of
104~km~s$^{-1}$, resulting in the difference $\Delta\!  V_1=4$~km~s$^{-1}$
between the stellar and \HI velocities. Assuming that locations and radial
velocities of the $N$ HVCs are uncorrelated and distributed uniformly within the
circle with the radius $R_0$ and the velocity interval $\Delta\! V_0$, the
probability $P$ to find {\it at least one} HVC located within $R_1$ from the
center of the Sculptor dSph and having radial velocity within $\Delta\! V_1$
from the velocity of the dwarf galaxy is given by the following expression:

\begin{equation} \label{eq1}
P=1-\left[1-\left(\frac{R_1}{R_0}\right)^2 \frac{2\,\Delta\! V_1}{\Delta\! V_0}\right]^{N+1}.
\end{equation}

\noindent (The exponent is $N+1$ because we also include the Sculptor cloud; the
velocity interval $\Delta\! V_1$ is multiplied by 2 because the difference in
velocities can be both positive and negative.) In our case, $\Delta\!
V_0=130$~km~s$^{-1}$, and the probability that the Sculptor cloud is an HVC is
$P=0.17$\%. Even if no velocity information was available and we had included
all the clouds in Table \ref{hvclistout} in our calculation, the probability
would still only be $P\sim 0.44$\%. If we considered the Sculptor clouds to be
two independent HVCs, the corresponding probability would be a few orders of
magnitude lower. Of course, either one or both of our assumptions (absence of
correlation, and uniform distribution) might prove to be wrong, so the
probability $P\sim 0.2$\% derived above should be treated with caution.

There is one fundamental flaw in the above probability derivations. To obtain
equation~(\ref{eq1}), we implicitly assumed that our choice of a dSph galaxy
(Sculptor) is an unbiased one. In reality, there are other Galactic dSphs, and
the reason we are so interested in Sculptor is because it happened to have \HI
gas in its vicinity with similar radial velocity. The correct question to ask
would be: ``What is the probability that at least one of the dSphs has at least
one HVC in its vicinity, which is located as close (or even closer) to the dwarf
galaxy (both spatially, and in radial velocity) as the Sculptor \ds{} cloud
is?'' To answer this question one would need to calculate corresponding
individual probabilities $P_i$ using equation~(\ref{eq1}) for the 8 known dSphs
(we exclude Sagittarius as it appears to be in the process of being disrupted by
the Galactic tidal field), and then to estimate the total probability as 

\begin{equation}
\label{eq2}
P_{tot}=1-\prod\limits_{i=1}^8 (1-P_i).
\end{equation}

\noindent Assuming for simplicity that the individual probabilities $P_i$ are
the same for all dSphs, and are equal to the derived above probability for the
Sculptor dSph ($P=0.17$\%), the total probability is
$P_{tot}=1-(1-P)^8\simeq1.4$\%.  Even in this more realistic approach, the
probability of the Sculptor clouds being HVCs appears to be very low.

It is clear from Figure~\ref{pksmom2} and Table~\ref{hvcparam} that the Sculptor
clouds are quite different from at least some of the other nearby features. NGC
300 is a spiral galaxy; its \HI cannot be confused with other structures in
Figure~\ref{pksmom2}.  The Milky Way's \HI in that direction, although beeing
near the South Galactic Pole line of sight, still has some leftover emission at
+40~\kms. This residual emission is relatively faint when compared with the bulk
emission at 0~\kms{} but is still very bright when compared with other
``background'' sources. A careful inspection of the dataset reveals that Region
2 is such an extension and it is most likely some sort of feature from the Milky
Way.

HVC 1625 has two distinct components: a large (1$\degr$) filament with a low
velocity dispersion and a compact (spatially unresolved) core with a higher
dispersion.  HVC 1640 is a compact object having a high dispersion, similar to
HVC 1625.  These objects have velocity structures that are very different from
the other clouds seen in Figure~\ref{pksmom2}.  Therefore they are probably of
different origin.

Region 1 is composed of two types of structure: a broad region with a low
dispersion and a very high dispersion compact core.  This apparently high
dispersion value is artificially inflated by the overlap of two clouds with
different velocities and should be treated as an artifact. The compact cloud is
probably an extension of the Galactic \hi{}, similar to the case of the Region
2. The other clouds of Region 1, HVC 1427 and HVC 1550 are all similar in size
and kinematic structure to the Sculptor clouds. It is conceivable then that
these clouds are part of a larger \HI stream. The shape of the SW cloud also
suggests that this link might be real. Analysis of the velocity field
(Figure~\ref{pksmom1}) suggests a velocity gradient between the SW cloud and HVC
1550. However, upon inspection of all HVCs (Table~\ref{hvclistin}
and~\ref{hvclistout}), there is no evidence for an \hi{} stream in this region.
The velocities of the nearby clouds are dispersed in an incoherent pattern.

One of the most interesting results obtained in the present work is the almost
perfect agreement between the \HI velocity for the both Sculptor clouds
(Table~\ref{tabspectre}) and the optical velocity of the dwarf
(Table~\ref{sclparam}).  The difference in velocities for the Parkes data is
$V_{\odot}^{opt}-V_{HI}=4\pm 3$~km~s$^{-1}$.  However, a kinematic agreement
between \HI clouds and \ds{} galaxies is to be regarded with caution. Alone,
this should never be considered as an evidence for an \HI association.

In the case of the Sculptor dSph, the fact that almost all of the detected \HI
emission is contained within the optical boundaries of the dwarf galaxy
(Figure~\ref{pksmom0}), the symmetric location of the two clouds relative to the
galactic center (Figure~\ref{camom0}), the closeness of the radial velocities
between the clouds and the dwarf, the low inferred probability $P_{tot}\simeq
2$\% for the clouds to be HVCs, and the presence of an arm coming out of the SW
cloud and pointing in the direction of the Sculptor dSph center
(Figure~\ref{camom0}), strongly favor the physical association of the clouds and
dwarf galaxy. However, in the absence of accurate distance measurements to the
\HI clouds, the HVC hypothesis remains a possible (though unlikely) alternative.

\subsection{Nature of the Clouds associated with the Sculptor \ds}
\label{nature}

Having argued that the NE and SW clouds are physically associated with the
Sculptor dSph, the important question to answer is whether the clouds are still
gravitationally bound to the dwarf galaxy, or if they have been removed by some
mechanism --- either internal (winds from red giants, supernovae type Ia), or
external (ram pressure stripping by the Galactic hot halo gas, Galactic tidal
field).

The symmetric location of the clouds relative to the center of the Sculptor dSph
along its proper motion vector (see Figure~\ref{camom0}) appears to be
consistent with the tidal removal picture. However, the orientation of the
stellar body of the dwarf which has a comparable angular extent to that of the
\HI gas (see Figure~\ref{pksmom0}) indicates that the tidal forces have not
played a major part in shaping the galaxy.  The biggest obstacle for another
external removal mechanism --- ram pressure stripping --- is the presence of the
NE cloud ahead of the proper motion (Figure~\ref{camom0}).

An important observational evidence for the clouds being gravitationally bound
to the dwarf would be a rotation signature for the gas.  C98 speculated on the
possibility of rotation of the NE and SW clouds around the center of the
Sculptor dwarf. It is clear from Figure~\ref{lv} that there is a velocity
gradient between the two clouds.  The classical Newtonian formula gives a
central mass of $\sim6.7\times 10^{5}$ \msol{} when using the Parkes velocity
information (rotation speed of 2.5 \kms{} at a distance of 20\arcmin, see
Table~\ref{tabspectre} and Figure~\ref{lv}). The ATCA velocity information gives
an enclosed mass of 2.7 $\times 10^{6}$ \msol{} with the same conditions but a
rotation speed of 5 \kms.  These estimates assume that the gas follow a circular
orbit viewed edge on, therefore no projection effect on the distance or velocity
of the clouds have been taken into account and the gas is considered to be
neither infalling nor expanding. \citet{mateo98} gives a total mass of 6.4
$\times 10^{6}$ \msol{} for the Sculptor dSph. The inferred rotation speed
values are comparable to or less than the internal velocity dispersion in the
clouds (see Table~\ref{hvcparam}).  In this respect, the Sculptor dSph would be
similar to low luminosity dIrr and dIrr/dSph galaxies, such as GR~8, Leo~A,
SagDIG, and LGS~3.

Our main argument against any \HI removal scenario (either internal or external)
is that the gas removal would not solve the ISM crisis in dSph galaxies
\citep{mateo98}. The alternative is that the NE and SW clouds are
gravitationally bound to the Sculptor dSph, and are part of its ISM.

MCB proposed a scenario which can explain the statistical differences between
the low luminosity dwarf galaxies in the Local Group. They showed that the FUV
radiation escaping from spiral galaxies can warm up and photoionize the ISM of
their dSph satellites, quenching star formation and making the ISM virtually
unobservable. Only during relatively short duration passages through the plane
of the host galaxy does the FUV radiation flux become small enough to allow the
ISM to recombine and potentially form stars. An important requirement of the
model is that many dSph galaxies should possess extended and very massive dark
matter halos (with virial temperature $\gtrsim 10^4$~K), allowing them to keep
the warm photoionized ISM with temperature $T\sim 10^4$~K gravitationally bound.
There is a growing number of evidence supporting the idea of the dSph galaxies
being more massive than the predictions of the mass-follows-light King model
\citep{odenkirchen01,kleyna01,hayashi02}.

The Sculptor dSph is located relatively close to the Milky Way ($<$ 80~kpc), and
is close to the southern Galactic pole. According to the photoevaporation model
of MCB, it is expected that the ionizing radiation escaping from the Galactic
disk should affect significantly the Sculptor \HI clouds.  Because the Sun is
located much closer to the Galactic center than to the Sculptor dSph, the
photoionized regions of the clouds would be seen approximately face-on.  In the
photoevaporation scenario, the photoionized gas expands with a speed comparable
to its sound speed ($\sim 10$~km~s$^{-1}$). First it moves normally to the \HI
cloud surface. If the \HII gas stays gravitationally bound to the galaxy, it
will soon decline from the normal direction. As a result, the H$_{\alpha}$
spectral line from the photoionized gas is expected to be blue-shifted by $\sim
5-10$~km~s$^{-1}$ relatively to the 21~cm spectral line from the neutral cloud.

\begin{figure}[tbh]
\begin{center}
\includegraphics[angle=270, width=0.45\textwidth]{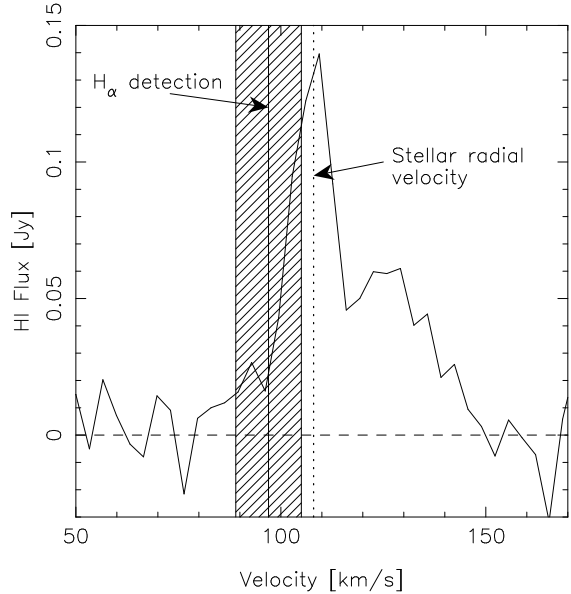}
\caption{H$_{\alpha}$ detection from the Sculptor dSph (\citealt{weiner2001}; B.
J. Weiner 2001, private communication).  Vertical filled stripe corresponds to
the radial velocity of the H$_{\alpha}$ emission from the southwest \HI cloud
inside a circle with a radius 5$\arcmin$ centered at $\alpha_{2000}=0^{\rm
h}58^{\rm m}22^{\rm s}$, $\delta_{2000}=-34\degr 00\arcmin$. The width of the
stripe reflects both formal and systematic errors. We also show the \HI spectral
line from the ATCA data integrated within the same circle area (solid line) and
the stellar radial velocity of the Sculptor dSph (vertical doted line).
\label{fig_halpha}}
\end{center}
\end{figure}

\citet{weiner2001} detected H$_{\alpha}$ emission from the SW cloud with a flux
of $220\pm 23$~mR.  The heliocentric radial velocity of the ionized gas is
$V_{HII}=97\pm 3\pm 5$~km~s$^{-1}$ (formal / systematic errors) (B. J. Weiner
2001, private communication). Figure~\ref{fig_halpha} shows both the
H$_{\alpha}$ detection and the \HI spectral line integrated over the same area.
The \HI line has a peak at $V_{HI}\simeq 109$~km~s$^{-1}$ (which is virtually
identical to the stellar radial velocity). The \HII gas appears to be
blue-shifted by $V_{HI}-V_{HII}=12\pm 8$~km~s$^{-1}$ relative to the \HI cloud,
which is consistent with the prediction of the photoevaporation model of MCB.
More accurate measurements of the radial velocity of the H$_{\alpha}$ emitting
gas are required to confirm this result with higher confidence.

\clearpage

Another interesting feature of Figure~\ref{fig_halpha} is the presence of a
red-shifted tail in the \HI spectral line. Both analytical \citep{bertoldi90}
and numerical \citep{lefloch94} calculations showed that under a wide range of
initial conditions photoevaporating interstellar clouds tend to a
quasi-equilibrium cometary state. The cometary tail in these models consists of
the neutral gas accelerated away from the cloud. The gas in the tail is neutral
because it is shielded from the ionizing radiation by the bulk of the cloud. The
red-shifted \HI gas in Figure~\ref{fig_halpha} could be a cometary tail of the
photoevaporating cloud.

\section{Conclusions}

New observations of the \HI gas in a large area of $7\degr\times 7\degr$ around
the Sculptor dSph are presented. Combination of single-dish (Parkes) and
interferometric (ATCA) observations allowed us to achieve both large angular
coverage and high angular resolution. These new data sets are of significantly
higher resolutions than any other previously released data sets in the line of
sight of the Sculptor \ds. Our principal results are:

\begin{enumerate}

\item Large angular coverage is required in order to claim to understand the \HI
properties of \ds{} galaxies. Our new data now shows the full extent of the C98
clouds. The size of these clouds is found to be much larger than the largest
angular scale ($\sim 22\arcmin$) the C98 observations were sensitive to.

\item The single-dish Parkes radiotelescope observations give a total \HI mass
for the C98 clouds of $2.3\times 10^5$~[D/79kpc]$^2$~M$_{\odot}$. The
heliocentric radial velocity of the gas is 104~km~s$^{-1}$.

\item Arguments that the C98 Sculptor clouds are physically associated with the
dwarf, thus at a distance of 79~kpc, are presented, the most important ones
being the closeness of the radial velocities for the \HI gas and the stars ($\Delta
V=4\pm 3$~km~s$^{-1}$ based on the Parkes data), and the location of almost all
of the \HI emission (88\% of the total flux) within the stellar body of the
dwarf (in projection). The unlikely possibility of the clouds being part of a
general ensemble of compact HVCs cannot be ruled out on the basis of the
available data.

\item The combination of the present \HI observations with the H$_\alpha$
emission line detection from the southwest cloud \citep{weiner2001} gives
support to the FUV radiation harassment model of MCB. The difference in radial
velocities between the neutral and ionized gas is found to be $12\pm
8$~km~s$^{-1}$, which is consistent with the Sculptor clouds being
photoevaporated by the Milky Way LyC radiation.

\end{enumerate}

We are grateful to Lister Staveley-Smith for helpfull discussions. We
acknowledge financial suport from NSERC, Canada and FQRNT, Qu\'ebec. The
Australia Telescope Compact Array and Parkes telescope are part of the Australia
Telescope which is funded by the Commonwealth of Australia for operation as a
National Facility managed by CSIRO.

\end{document}